\documentclass{article}

\pdfoutput=1

\usepackage{arxiv}

\usepackage{amsmath,amssymb}
\usepackage{algorithmic}
\usepackage{textcomp}
\usepackage{forest}
\usepackage{textgreek}
\usepackage{multicol}
\usepackage{multirow}
\usepackage{subcaption}

\usepackage[utf8]{inputenc} 
\usepackage[T1]{fontenc}    
\usepackage{hyperref}       
\usepackage{url}            
\usepackage{booktabs}       
\usepackage{amsfonts}       
\usepackage{nicefrac}       
\usepackage{microtype}      
\usepackage{cleveref}       
\usepackage{lipsum}         
\usepackage{graphicx}
\usepackage[numbers]{natbib}
\usepackage{doi}

\title{Ovarian Cancer Diagnostics using Wavelet Packet Scaling Descriptors}

\newif\ifuniqueAffiliation
\uniqueAffiliationtrue

\ifuniqueAffiliation 
\author{ \href{https://orcid.org/0009-0000-1733-1613}{\includegraphics[scale=0.03]{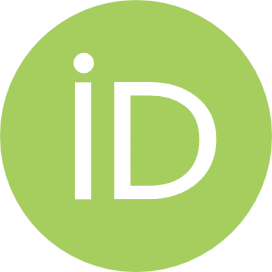}\hspace{1mm}Raymond J.~Hinton, Jr.} \\
	Department of Statistics\\
	Texas A\&M University\\
	College Station, TX 77843 \\
	\texttt{rhinton@tamu.edu} \\
	\And
	\href{https://orcid.org/0000-0002-2342-9825}{\includegraphics[scale=0.03]{orcid.png}\hspace{1mm}Jihyun~Byun} \\
	Department of Statistics\\
	Texas A\&M University\\
	College Station, TX 77843 \\
	\texttt{jhbyun@tamu.edu} \\
	\AND
	\href{https://orcid.org/0000-0001-6794-4776}{\includegraphics[scale=0.03]{orcid.png}\hspace{1mm}Dixon~Vimalajeewa} \\
	Department of Statistics \\
	University of Nebraska - Lincoln \\
    Lincoln, NE 68583 \\
	\texttt{hvimalajeewa2@unl.edu} \\
	\And
	\href{https://orcid.org/0000-0001-9155-9325}{\includegraphics[scale=0.03]{orcid.png}\hspace{1mm}Brani~Vidakovic} \\
	Department of Statistics\\
	Texas A\&M University\\
	College Station, TX 77843 \\
	\texttt{brani@stat.tamu.edu} \\
}
\else
\usepackage{authblk}

\newbox{\orcid}\sbox{\orcid}{\includegraphics[scale=0.03]{orcid.png}} 
\author[1]{%
	\href{https://orcid.org/0000-0000-0000-0000}{\usebox{\orcid}\hspace{1mm}David S.~Hippocampus\thanks{\texttt{hippo@cs.cranberry-lemon.edu}}}%
}
\author[1,2]{%
	\href{https://orcid.org/0000-0000-0000-0000}{\usebox{\orcid}\hspace{1mm}Elias D.~Striatum\thanks{\texttt{stariate@ee.mount-sheikh.edu}}}%
}
\affil[1]{Department of Computer Science, Cranberry-Lemon University, Pittsburgh, PA 15213}
\affil[2]{Department of Electrical Engineering, Mount-Sheikh University, Santa Narimana, Levand}
\fi

\renewcommand{\headeright}{R.J. Hinton et al.}
\renewcommand{\undertitle}{}
\renewcommand{\shorttitle}{Ovarian Cancer Diagnostics using Wavelet Packet Scaling Descriptors}

\hypersetup{
pdftitle={Ovarian Cancer Diagnostics using Wavelet Packet Scaling Descriptors},
pdfsubject={stat.AP},
pdfauthor={Raymond J. Hinton, Jr., Jihyun Byun, Dixon Vimalajeewa, Brani Vidakovic},
pdfkeywords={self-similarity, wavelet transform, wavelet packets, classification, ovarian cancer},
}

\begin{document}
\maketitle

\begin{abstract}
Detecting early-stage ovarian cancer accurately and efficiently is crucial for timely treatment. Various methods for early diagnosis have been explored, including a focus on features derived from protein mass spectra, but these tend to overlook the complex interplay across protein expression levels. We propose an innovative method to automate the search for diagnostic features in these spectra by analyzing their inherent scaling characteristics. We compare two techniques for estimating the self-similarity in a signal using the scaling behavior of its wavelet packet decomposition. The methods are applied to the mass spectra using a rolling window approach, yielding a collection of self-similarity indexes that capture protein interactions, potentially indicative of ovarian cancer. Then, the most discriminatory scaling descriptors from this collection are selected for use in classification algorithms. To assess their effectiveness for early diagnosis of ovarian cancer, the techniques are applied to two datasets from the American National Cancer Institute. Comparative evaluation against an existing wavelet-based method shows that one wavelet packet-based technique led to improved diagnostic performance for one of the analyzed datasets (95.67\% vs. 96.78\% test accuracy, respectively). This highlights the potential of wavelet packet-based methods to capture novel diagnostic information related to ovarian cancer. This innovative approach offers promise for better early detection and improved patient outcomes in ovarian cancer.
\end{abstract}

\keywords{self-similarity \and wavelet transform \and wavelet packets \and classification \and ovarian cancer}

\section{Introduction}
According to the American Cancer Society, ovarian cancer is expected to cause 13,270 deaths in 2023, making it the fifth leading cause of cancer-related deaths among women in the United States. The lifetime risk of a woman developing ovarian cancer is approximately 1 in 78, while her risk of dying from the disease is about 1 in 108. Early detection is crucial in reducing mortality risk, as it has a five-year survival rate of around 94\% \cite{ACS2022}. However, most cases of ovarian cancer are diagnosed at a more advanced stage because early symptoms may not be noticeable and can resemble other less serious conditions. Additionally, preventative screening has not been effective in prospective randomized controlled trials. Only 18\% of ovarian cancer diagnoses are made at an early stage (Localized/Stage I), and the five-year survival rates for advanced-stage diagnoses range from 72\% (Regional/Stage II-III) to 32\% (Distant/Stage IV) \cite{SEER2023}. Therefore, the development of early-detection techniques for ovarian cancer is of utmost importance.

Proteomic profiles used to detect ovarian cancer typically involve the analysis of a wide range of proteins in biological samples, such as serum, plasma, or tissue, to identify patterns or signatures that are indicative of the presence or progression of ovarian cancer. These profiles can be generated using various proteomic techniques, including mass spectrometry, immunoassays, and protein microarrays. Mass spectrometry is the most frequently used technique among these techniques \cite{Ghose2022proteomics}. For instance, MALDI-TOF MS (Matrix-Assisted Laser Desorption/Ionization Time-of-Flight Mass Spectrometry) is utilized to analyze serum or tissue samples for protein profiles associated with ovarian cancer \cite{Li2020performance}. It allows for the detection of specific protein peaks or patterns that may be associated with the disease. Existing studies primarily involve identifying the expression of a combination of proteins or protein patterns that can distinguish between ovarian cancer patients and healthy individuals or those with other conditions. The goal is to improve the accuracy and specificity of ovarian cancer detection and potentially aid in early diagnosis. 

Data analysis and bioinformatics tools play a critical role in ovarian cancer diagnostics and research. They help researchers and clinicians process and interpret large datasets to identify potential biomarkers, understand disease mechanisms, and make diagnostic or therapeutic decisions. Among the commonly used data analysis and bioinformatics tools for ovarian cancer diagnostics, machine learning tools such as predictive models and classifiers have become more popular with the growing interest in involvement of modern data science and artificial intelligence. For instance, \cite{Zhiqiao2023} proposes a novel approach for identifying prognostic signatures in ovarian cancer patients using deep learning to identify prognostic gene signatures using gene expression data from multiple public databases. A machine learning model integrated into a nanosensor is proposed to detect ovarian cancer from a blood test in \cite{Macklin2020}, while a distance novel variance-based modality is proposed to detect ovarian cancer from protein mass spectra in \cite{Dixon2023}. References \cite{Chakraborty2023} and \cite{Liberto2022} review existing and potential biomarkers for use in ovarian cancer diagnostics and also discuss a range of opportunities for developing novel methods. 

While data analysis and bioinformatics tools have significantly advanced our ability to study and diagnose ovarian cancer, they are not without their drawbacks and limitations. Some of the limitations of existing tools used to detect ovarian cancer include interpretation challenges, biological noise, standardization, clinical relevance, and dynamic nature of data. For instance, \cite{Liberto2022} provides an overview of the current challenges associated with the existing methods. Therefore, it is important to note that ongoing research in proteomics continues to refine these approaches and discover new protein biomarkers for ovarian cancer.

Wavelets are mathematical functions that decompose a signal at different frequencies with a temporal resolution that varies according to the frequency scale. This can be useful for, broadly, uncovering complex patterns occurring at different frequencies in a signal. Wavelet-based data analysis has been used in many fields, including the processing and analysis of biomedical data in bioinformatics \cite{Dixon2022, Dixon2023a, Dixon2023c, Dixon2023d}. In the context of ovarian cancer research, examples of wavelet-based techniques and tools that have been used include feature extraction, such as in \cite{Alqudah2019ovarian} and \cite{Dixon2023}, where new predictor variables are created based on the data that are then used in classification studies. Reference \cite{Maria2022Denoising} applied denoising methods to low-dose computed tomography (LDCT) images to improve the quality of information obtained at safer levels of radiation. In \cite{Jang2019multiresolution}, a wavelet-based multiresolution analysis is used to correct biases in whole-genome sequencing for identifying ovarian cancer-related genetic variations. Wavelets offer several advantages for analyzing biomedical data, such as the ability to summarize complex interplay between different frequency components in a signal which can lead to simpler models and thus improve interpretability. Wavelets are also a flexible tool since different wavelet families and sizes can be chosen, and different levels of analysis can be performed depending on the size of the signal and wavelet. Overall, wavelet-based approaches can be valuable in characterizing the complexity of biological systems relevant to ovarian cancer.

Wavelet packets produce a richer representation of the original signal by extending the concept of the standard wavelet transform. By additionally analyzing the detail (or residual) coefficients at the end of each step, rather than only the signal approximation coefficients, wavelet packets can provide better characterizations of certain types of signals, such as those with significant high-frequency components. Wavelet packets have been used in applications such as signal compression \cite{Coifman1994signal}, bearing fault diagnosis \cite{Nikolaou2002rolling}, analyzing variations in Earth’s magnetic field \cite{Mandrikova2013analysis}, and a wide variety of other areas. In medicine, wavelet packets have particularly been applied to analyzing electroencephalographic (EEG) signals \cite{Wang2021application} and also for medical imaging such as ultrasounds \cite{Cincotti2001ultrasound}. Each level of a wavelet packet decomposition is the same length as the original signal, meaning that the representation is redundant. This leads to a choice of how to efficiently recover or otherwise summarize the original signal. 

In this study, we compare two wavelet packet-based techniques, \cite{Wang2006} and \cite{Jones1996wavelet}, for generating features from blood protein mass spectra and use these features in classification algorithms to detect ovarian cancer. The potential of these techniques to outperform traditional wavelet-based methods is also demonstrated, highlighting their efficacy for this task. A broader contribution of our paper is the use of wavelet packets in the context of scaling. There are relatively few references that employ wavelet packets to estimate Hurst exponents and other scaling indexes, compared to references of their use in the regression, signal representation, and denoising contexts. This is especially true for signals of sound, where wavelet packets produce the most parsimonious representations \cite{Wickerhauser1996adapted}.

The remainder of the paper is organized as follows: Section \ref{sec:moti_stdy} gives an overview of the motivating study and datasets used for the analysis. The techniques used in this study, fundamentals of wavelet transform, and background for measuring self-similarity in signals using wavelet transforms are presented in Section \ref{sec:Methods}. Section \ref{sec:Classification} provides data analysis procedures and results, followed by a discussion in Section \ref{sec:Discussion}. Concluding remarks are in Section \ref{sec:Conclusion}.

\section{Motivating Study}\label{sec:moti_stdy}
\begin{figure}[!t]
\centering
 \includegraphics[width=\linewidth]{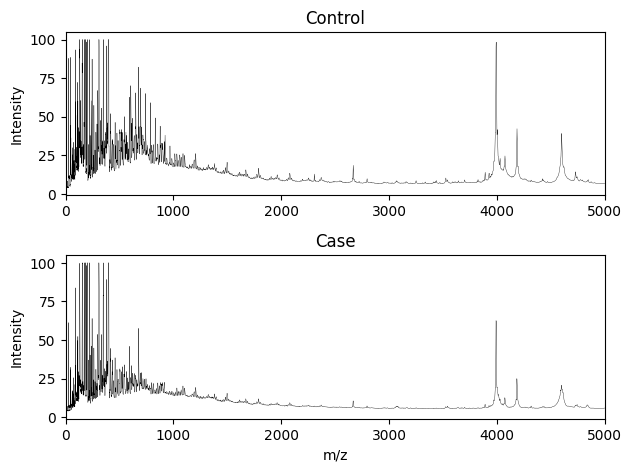}
        \caption{Examples of protein mass spectra from the Control and Case groups, zoomed in to 0-5000 m/z to show detail. Note that the Case spectra exhibits suppressed peaks in the 4000-5000 m/z region relative to the Control, while both samples show complex differences in the 0-1000 m/z region. This demonstrates the need for diagnostic techniques that look beyond individual peak intensities.}
        \label{fig-0}
\end{figure}

This study examines two ovarian cancer datasets obtained from the American National Cancer Institute internet repository \cite{R39}. These datasets comprise protein mass spectra derived from the surface-enhanced laser desorption-ionization time-of-flight (SELDI-TOF) mass spectrometer, utilizing blood samples collected from both ovarian cancer patients and healthy individuals. The first dataset, referred to as {\it Ovarian} 4-3-02, was gathered using the weak cation exchange (WCX2) protein chip and the PBI SELDI-TOF mass spectrometer. It consists of 216 protein mass spectra, encompassing 100 cancer samples, 100 non-cancer samples, and 16 benign samples. The second dataset, known as {\it Ovarian} 8-7-02, was also collected using the WCX2 protein chip, but with an upgraded version of the PBI SELDI-TOF mass spectrometer. Consequently, {\it Ovarian} 8-7-02 differs significantly from {\it Ovarian} 4-3-02 in terms of the method used to record the protein mass spectra. This dataset comprises protein mass spectra collected from 162 women diagnosed with cancer and 91 women without a cancer diagnosis. In both datasets, the protein mass spectra are represented by the intensities of 15,153 peptides, defined by their mass-to-charge ratio (m/z). Fig. \ref{fig-0} displays two sample protein mass spectra, one from the healthy (Control) group and the other from the cancer (Case) group, selected from the {\it Ovarian} 4-3-02 dataset. The $x$-axis represents the m/z values, while the $y$-axis represents the spectral intensity. Additional information about these datasets and the related studies can be found in \cite{R39,R2}.

\section{Methods}\label{sec:Methods}
This Section introduces the techniques used to extract discriminatory features by assessing self-similar properties of protein mass spectra in the wavelet domain. First, we provide a brief overview of wavelet transforms, wavelet packets, and the notion of self-similarity. Then, we describe two wavelet packet-based methods for assessing self-similarity. Appendices \ref{sec:wt1_Appendix}-\ref{sec:wang_Appendix} provide technical details about these techniques.

\subsection{Wavelet Transform}\label{sec:wt}
Wavelet transforms (WTs) are extensively utilized tools in the field of signal processing. When applied to a signal, they break down the signal into localized contributions in both time and frequency/scale, resulting in a hierarchical representation. This representation enables simultaneous analysis at various resolutions or scales, facilitating the exploration of signal properties that may not be evident in the domain of the original data (acquisition domain).

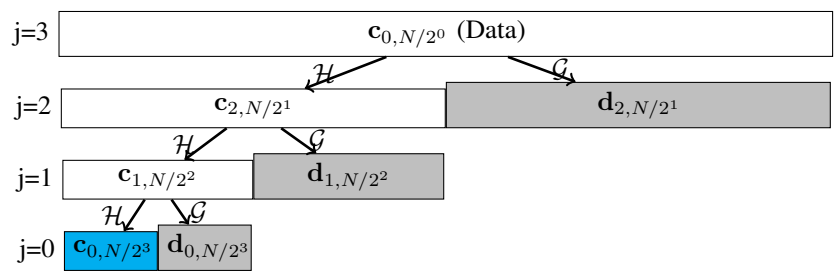
\begin{figure*}[!t]
\center
  \begin{forest}
    for tree = {grow=south,
      l sep=0.0cm,
      s sep=0.0cm,
      minimum height=0.5cm,
      minimum width=1cm,
      edge={->,line width=1pt},
      text centered,
      }
    [$\mathbf{c}_{0,N/2^0}$ (Data), draw, text width=10.05cm, label={left:{j=3}}
      [$\mathbf{c}_{2,N/2^1}$, draw, text width=4.87cm, edge label={node[midway,left]{$\mathcal{H}$}}, label={left:{j=2}}
       [$\mathbf{c}_{1,N/2^2}$, draw, text width=2.29cm, edge label={node[midway,left]{$\mathcal{H}$}}, label={left:{j=1}}
         [$\mathbf{c}_{0,N/2^3}$, draw, text width=1cm, fill=cyan, edge label={node[midway,left]{$\mathcal{H}$}}, label={left:{j=0}}]
         [$\mathbf{d}_{0,N/2^3}$, draw, text width=1cm, fill=lightgray, edge label={node[midway,right]{$\mathcal{G}$}}]
       ]
       [$\mathbf{d}_{1,N/2^2}$, draw, text width=2.29cm, fill=lightgray, edge label={node[midway,right]{$\mathcal{G}$}}
       ]
      ]
      [$\mathbf{d}_{2,N/2^1}$, draw, text width=4.87cm, fill=lightgray, edge label={node[midway,right]{$\mathcal{G}$}}
        ]
      ]
    ]
  \end{forest}
  \caption{Illustration of three levels of decomposition ($j = 2, 1, 0$) of the discrete wavelet transform of a signal of length $N = 2^J (J \geq 3)$. At each decomposition level $j$, $\mathbf{c}_{j, N/2^{J-j}}$ denotes smoothing coefficients while $\mathbf{d}_{j, N/2^{J-j}}$ denotes the detail wavelet coefficients. The second index for each vector indicates their length, to emphasize that the number of detail coefficients is halved at each level.}
  \label{fig:DWT_Method}
\end{figure*}

The discrete wavelet transform (DWT), a popular variant of wavelet transforms, has become a standard tool for analyzing complex data signals, particularly in domains where discrete data is examined. DWTs are linear transforms that can be represented using orthogonal matrices. However, the computational cost of performing the DWT in matrix form increases as the length of the signal grows. To address this, a fast algorithm based on filtering was proposed by Mallat for computational efficiency \cite{Mallat1989}. This algorithm involves a series of successive convolutions utilizing a wavelet-specific low-pass filter and its mirrored high-pass filter. Operators $\mathcal{H}$ and $\mathcal{G}$ can be constructed with each filter by combining them with a decimation step (retaining every second coefficient of the convolution). By repeatedly applying these operators to the signal (and then the resulting "approximation" coefficients after the first level of decomposition), a multiresolution representation of the signal is generated. The resulting representation comprises a smooth approximation and a hierarchy of detail coefficients at different resolutions (or scale indexes) and locations within the same resolution, as show in Fig. \ref{fig:DWT_Method}. For detailed instructions on calculating these coefficients, please refer to Appendix \ref{sec:wt1_Appendix}. These coefficients provide a description of the signal at different scales and locations, serving as a valuable means of characterizing the signal's multi-scale dynamics.

\subsection{Wavelet Packets}\label{sec:wp}

The wavelet packet decomposition (WPD) is an extension of the discrete wavelet transform that yields a richer (yet redundant) representation of the original signal, which can be arranged in a "table" or binary tree with nodes containing coefficients. The extended analysis is potentially able to capture certain patterns better than the DWT. However, a choice must be made as to what representation from all those produced by the WPD best characterizes the signal.

\begin{figure*}[!t]
\center
  \begin{forest}
    for tree = {grow=south,
      l sep=0cm,
      s sep=0cm,
      minimum height=0.5cm,
      minimum width=1cm,
      edge={->,line width=1pt},
      text centered
      }
    [$\mathbf{w}_{3,0}$ (Data), draw, text width=10.05cm, label={left:{j=3}}
      [$\mathbf{w}_{2,0}$, draw, text width=4.87cm,  edge label={node[midway,left]{$\mathcal{H}$}}, label={left:{j=2}}
       [$\mathbf{w}_{1,0}$, draw, text width=2.29cm, edge label={node[midway,left]{$\mathcal{H}$}}, label={left:{j=1}}
         [$\mathbf{w}_{0,0}$, draw, text width=1cm, edge label={node[midway,left]{$\mathcal{H}$}}, fill=cyan, label={left:{j=0}}]
         [$\mathbf{w}_{0,1}$, draw, text width=1cm, edge label={node[midway,right]{$\mathcal{G}$}}, fill=lightgray]
       ]
       [$\mathbf{w}_{1,1}$, draw, text width=2.29cm, edge label={node[midway,right]{$\mathcal{G}$}}, fill=lightgray
         [$\mathbf{w}_{0,2}$, draw, text width=1cm]
         [$\mathbf{w}_{0,3}$, draw, text width=1cm, edge label={node[midway,right]{$\mathcal{G}$}}]
       ]
      ]
      [$\mathbf{w}_{2,1}$, draw, text width=4.87cm, edge label={node[midway,right]{$\mathcal{G}$}}, fill=lightgray
        [$\mathbf{w}_{1,2}$, draw, text width=2.29cm
          [$\mathbf{w}_{0,4}$, draw, text width=1cm]
          [$\mathbf{w}_{0,5}$, draw, text width=1cm]
        ]
        [$\mathbf{w}_{1,3}$, draw, text width=2.29cm
          [$\mathbf{w}_{0,6}$, draw, text width=1cm]
          [$\mathbf{w}_{0,7}$, draw, text width=1cm]
        ]
      ]
    ]
  \end{forest}
  \caption{Diagram of the coefficient nodes of a wavelet packet tree. The highlighted nodes represent the DWT, showing that the DWT is a subset of the WPD. In this diagram, the second index indicates the number of the node within the level.}
  \label{fig:WPT_Methods}
\end{figure*}
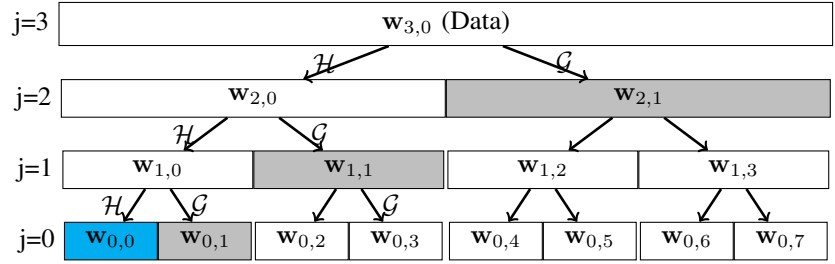

The wavelet packet table is created by applying the low- and high-pass operators $\mathcal{H}$ and $\mathcal{G}$ to both the approximation and detail coefficients, not just the approximation coefficients as in the DWT. This creates $2^{J-j}$ sets, or nodes, of coefficients at each level $j$, where the original signal is considered to be level $J$, the first level of decomposition is level $J-1$, etc. The result is that each level of the wavelet packet table has $N = 2^J$ coefficients, whereas the number of coefficients in the DWT is reduced by half at each level, i.e. $N/2^{J-j}$ (except for the final level which contains the approximation coefficients). Fig. \ref{fig:WPT_Methods} shows an example of a wavelet packet tree. The gray nodes of the tree correspond to the standard DWT, showing that it is a special case of the WPD. The coefficients in the vector $\mathbf{w}_{0, 0}$ (the DWT approximation coefficients after three levels of decomposition) are the result of the operator combination $\mathcal{HHH}$. As a further example, vector $\mathbf{w}_{0, 3}$ is a result of the operator combination $\mathcal{GGH}$ (in this mathematical notation, the first operator is the last one that is applied). More details about calculating the WPD can be found in Appendix \ref{sec:wpt_Appendix}.

The additional analysis with high-pass filters offers the potential to better characterize high-frequency components in the original signal. However, since each node in the tree contains $N/2^{J-j}$ coefficients at level $j$ (for an original signal of length $N$), the wavelet packet table is redundant in the sense that each level of the tree contains the same number of coefficients as in the original signal. Therefore, we must use some different method when compared to the DWT to extract information from these coefficients. In the following sections, we describe two such methods – one from Wang {\it et al.} \cite{Wang2006} that estimates the Hurst exponent using a spectral slope similar to that for the DWT, and another from Jones {\it et al.} \cite{Jones1996wavelet} that chooses the best basis representation from the wavelet packet table and estimates the Hurst exponent using power law scaling of the best basis coefficients.

\subsection{Self-similarity}\label{sec:ss}
\begin{figure}[!t]
\centering
    \includegraphics[width=\linewidth]{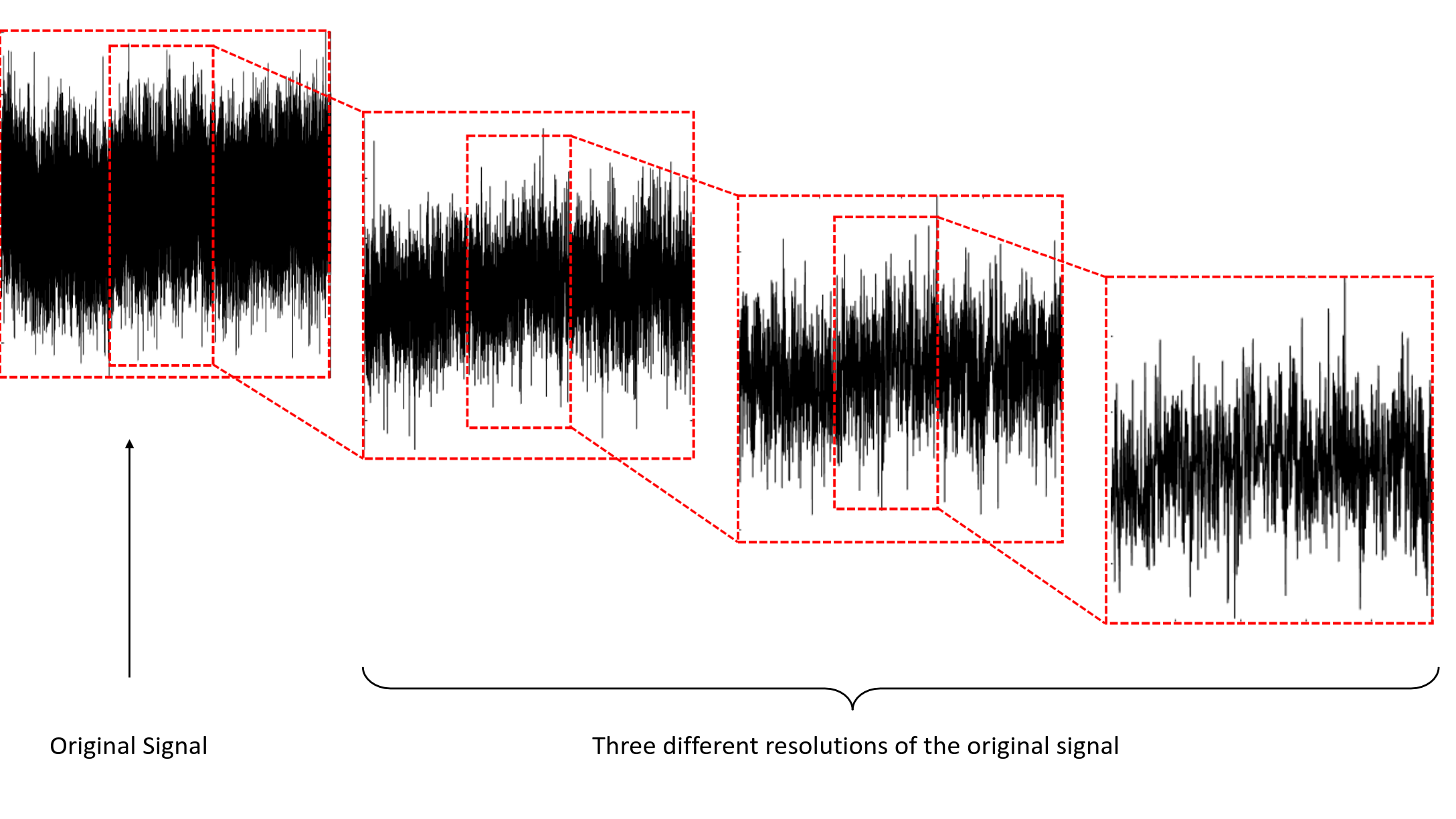}
    \caption{Example of a high-frequency signal at three different resolutions. The signal exhibits similar properties and behaviors (e.g. mean, variance) when examined at different resolutions. This quality is known as the self-similar nature of the signal.}
    \label{fig-00a}
\end{figure}

Self-similarity is an omnipresent phenomenon that characterizes high frequency time series obtained in different contexts: health, geoscience, economics, and physics, to list a few. This behavior refers to the stochastic similarity in a signal when viewed at different scales, an example of which is show in Fig. \ref{fig-00a}. In the wavelet domain, self-similarity is quantified commonly by using the wavelet spectra. The wavelet spectrum of a signal consists of wavelet log energies (logarithm of the average of squared wavelet detail coefficients) as a function of a resolution level. 

In general, it is accepted that a given signal possesses the self-similar property if its wavelet spectrum exhibits a regular decay with an increase of resolution scales. The rate of the log energy decay (slope) is connected with the Hurst exponent ($H \in [0, 1]$), which characterizes the regularity of the signal. To estimate the slope using the standard DWT, the log energy for each level is calculated with corresponding wavelet detail coefficients. Then, the linear regression of log energies to the scale index is found, and $H$ is computed as $H = -(slope +1)/2$. The spectral slope theoretically ranges from $-3$ to $-1$. Larger slopes ($>-2$, i.e., shallower) indicate a higher degree of persistence (i.e., more regular/smooth signal), while smaller slopes ($<-2$, i.e., steeper) indicate a higher degree of anti-persistency and intermittency. Thus, the spectral slope serves as a measure of signal regularity, since signals with a Hurst exponent close to $1$ are more regular (smooth), while signals with a small Hurst exponent are highly irregular. Overall, the degree of regularity expressed by either spectral slope, or equivalently by the Hurst exponent, represents an informative summary of a complex and noisy signal for which standard statistical summaries (moments, trends, etc.) may be irrelevant.

\subsection{Wang et al. method}\label{sec:Wang}
The \cite{Wang2006} method is similar to the DWT-based method in that it estimates a spectral slope for a signal using log energies. The difference is that the former uses the richer wavelet packet decomposition, and thus utilizes more detail coefficients at each level.

\begin{figure*}[!t]
\center
  \begin{forest}
    for tree = {grow=south,
      l sep=0cm,
      s sep=0cm,
      minimum height=0.5cm,
      minimum width=1cm,
      edge={->,line width=1pt},
      text centered
      }
    [$\mathbf{w}_{3,0}$ (Data), draw, text width=10.05cm, label={left:{j=3}}
      [$\mathbf{w}_{2,0}$, draw, text width=4.87cm,  edge label={node[midway,left]{$\mathcal{H}$}}, label={left:{j=2}}
       [$\mathbf{w}_{1,0}$, draw, text width=2.29cm, label={left:{j=1}}
         [$\mathbf{w}_{0,0}$, draw, text width=1cm, label={left:{j=0}}]
         [$\mathbf{w}_{0,1}$, draw, text width=1cm, fill=lightgray, edge label={node[midway,right]{$\mathcal{G}$}}]
       ]
       [$\mathbf{w}_{1,1}$, draw, text width=2.29cm, fill=lightgray, edge label={node[midway,right]{$\mathcal{G}$}}
         [$\mathbf{w}_{0,2}$, draw, text width=1cm]
         [$\mathbf{w}_{0,3}$, draw, text width=1cm, fill=lightgray, edge label={node[midway,right]{$\mathcal{G}$}}]
       ]
      ]
      [$\mathbf{w}_{2,1}$, draw, text width=4.87cm, fill=lightgray,  edge label={node[midway,right]{$\mathcal{G}$}}
        [$\mathbf{w}_{1,2}$, draw, text width=2.29cm, 
          [$\mathbf{w}_{0,4}$, draw, text width=1cm]
          [$\mathbf{w}_{0,5}$, draw, text width=1cm, fill=lightgray, edge label={node[midway,right]{$\mathcal{G}$}}]
        ]
        [$\mathbf{w}_{1,3}$, draw, text width=2.29cm, fill=lightgray, edge label={node[midway,right]{$\mathcal{G}$}}
          [$\mathbf{w}_{0,6}$, draw, text width=1cm]
          [$\mathbf{w}_{0,7}$, draw, text width=1cm, fill=lightgray, edge label={node[midway,right]{$\mathcal{G}$}}]
        ]
      ]
    ]
  \end{forest}
  \caption{Diagram of a wavelet packet tree with detail nodes highlighted. These nodes result from applying a high-pass operator, $\mathcal{G}$, to their parent nodes, as indicated with the labels. These are the nodes used for calculating level-wise energies in Wang {\it et al.}'s method.}
  \label{fig:WPT_Wang}
\end{figure*}
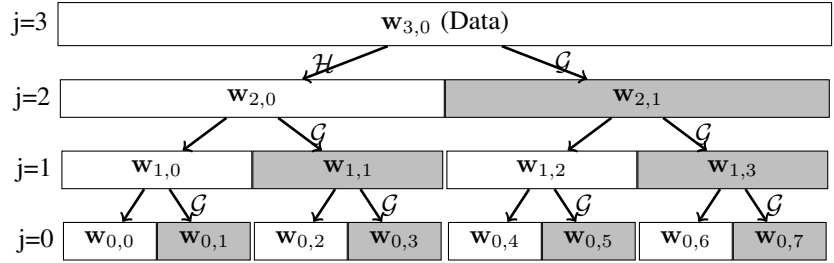

As described in Section \ref{sec:wp}, each level $j$ of a wavelet packet tree contains $2^{J-j}$ nodes - half containing detail coefficients and the other half containing approximation coefficients - that are the results of applying high-pass and low-pass filters, respectively, to the coefficients in the parent nodes. Thus, after the first level, each level contains multiple nodes with detail coefficients, and the Wang {\it et al.} method uses all such detail nodes to calculate an energy for the level. In the operator notation, these are any nodes that result from an operation beginning with $\mathcal{G}$, which are the shaded nodes in Fig. \ref{fig:WPT_Wang}. The energy for individual detail nodes is calculated similarly to the DWT-based method, and the energy for each level is just the average of the detail node energies. This results in an energy for each level, with which the wavelet spectra can be plotted and the spectral slope can be estimated in the same way as the DWT method. The Hurst exponent is estimated using $H = -(slope/2)$, as stated in \cite{Wang2006}.

Wang {\it et al.} showed that this method produced estimates of the Hurst exponent with lower absolute error for simulated fractal Brownian motion (fBm) signals at most values of $H$ when compared to DWT and other methods. We performed similar experiments using the DWT and Wang {\it et al.} methods that included calculating the standard deviation of the Hurst exponent estimates. Specifically, $n = 1000$ fBm signals of length $N = 2^{10} = 1024$ were simulated for several values of $H$. The full wavelet packet tree with $j_0 = \log_2 N = 10$ levels was calculated with the Haar wavelet. The DWT wavelet spectra was calculated using the first detail nodes in each level, and the Wang {\it et al.} method wavelet spectra was calculated as described above. The Hurst exponents were calculated for each method as described above, and the averages and standard deviations of the estimates across the simulations are shown in Table \ref{tab:wavelet_comparison}. The bias of the Wang {\it et al.} estimator is lower than that for the DWT-based method in five out of nine of the tested $H$ values, and the standard deviation of the $H$ estimates is lower for the Wang {\it et al.} method for all $H$ values. Simulations with the Jones {\it et al.} method (described in the next section) showed some of the lowest overall standard deviations. However, most of these were coupled with the highest overall bias. Therefore, among these three methods, the Wang {\it et al.} method may be a good compromise between bias and variance. 

\begin{table}[!t]
\centering
\begin{tabular}{|c|c|c|c|}
\hline
\multirow{2}{*}{{\bf H}} & {\bf DWT} & {\bf Wang et al.} & {\bf Jones et al.} \\
 & ({\it mean $\pm$ st. dev.}) & ({\it mean $\pm$ st. dev.}) &  ({\it mean $\pm$ st. dev.})  \\
\hline
0.1 & -0.0303 $\pm$ 0.0970 & \textbf{0.1049} $\pm$ \textbf{0.0248} &  0.1306 $\pm$ 0.0352 \\
0.2 & 0.1211 $\pm$ 0.1001 & \textbf{0.2071} $\pm$ \textbf{0.0385} & 0.2171 $\pm$ 0.0426  \\
0.3 & 0.2498 $\pm$ 0.1017  & 0.3103 $\pm$\textbf{ 0.0488} & \textbf{0.3086} $\pm$ 0.0510  \\
0.4 & 0.3712 $\pm$ 0.1031 & 0.4148 $\pm$ 0.0577 & \textbf{0.3930} $\pm$ \textbf{0.0519} \\
0.5 & \textbf{0.4870} $\pm$ 0.1091 & 0.5202 $\pm$ 0.0663 & 0.4460 $\pm$ \textbf{0.0461}  \\
0.6 & \textbf{0.6000} $\pm$ 0.1126 & 0.6256 $\pm$ 0.0748 & 0.5176 $\pm$ \textbf{0.0477} \\
0.7 & \textbf{0.7079} $\pm$ 0.1180 & 0.7288 $\pm$ 0.0822 & 0.6312 $\pm$ \textbf{0.0601}  \\
0.8 & \textbf{0.8068} $\pm$ 0.1217 & 0.8251 $\pm$ 0.0857 & 0.7603 $\pm$ \textbf{0.0769}\\
0.9 & 0.8922 $\pm$ 0.1125 & \textbf{0.9069} $\pm$ \textbf{0.0818} & 0.9080 $\pm$ 0.1016\\
\hline
\end{tabular}
\caption{Simulation results from estimating the Hurst exponent for known fractal Brownian motion (fBm) signals. The closest average estimates and lowest standard deviation for each $H$ are in bold. The Haar wavelet was used with the DWT-based and Wang {\it et al.} methods, and the SYMM4 wavelet was used with the Jones {\it et al.} method. Ten levels of WPD were performed for the DWT-based and \cite{Wang2006} methods, and nine were performed for the \cite{Jones1996wavelet} method.}
\label{tab:wavelet_comparison}
\end{table}

\subsection{Jones et al. method}\label{sec:Jones}

The method by \cite{Jones1996wavelet} uses a specially chosen subset of the wavelet packet coefficients to estimate the Hurst exponent according to a power law scaling relationship.

There are many valid bases contained in the wavelet packet library, so it is desirable to define some criteria with which to choose the best basis. Any particular basis may contain wavelet packets which do not match the original signal well, but are otherwise not entirely negligible. We could aim to concentrate the amount of information described with well-matching packets, and minimize the information described with poorly-matching packets. Thus, we seek to minimize an \emph{information cost functional} that is "large when coefficients are roughly the same size and small when all but a few coefficients are negligible" \cite{Coifman1992entropy}. Entropy is an example of such a cost functional and is the one used in the method by Jones {\it et al.} The best basis is constructed by selecting a subset of nodes which overall minimize the entropy (while still forming an orthonormal basis). There are a large number of potential bases, but the algorithm proposed by \cite{Coifman1992entropy} has a complexity that is proportional only to the number of nodes in the tree. Further details about selecting the best basis can be found in Appendix \ref{sec:bb_Appendix}.

Jones {\it et al.} propose that best basis wavelet packet coefficients can be described with a power law relationship – Korcak's Law – which measures the scaling behavior of a quantity. More information about the law can be found in \cite{Mandelbrot1982fractal} and a history in \cite{Imre2016fractals}. Specifically, the law relates the number of objects that are larger than some threshold size to that threshold size. In the context of wavelet packet coefficients, Jones {\it et al.} propose using
\begin{equation}\label{korcakslaw}
    N_r (A > a) \approx k a^{- \delta} \text{,}
\end{equation}
with the sorted absolute coefficient values as $a$, an index over the sorted absolute coefficients as $N_r$, and $(1+H)$ as $\delta$, the slope exponent. Then, a log-log plot of the sorted absolute coefficient values versus the index can be used to estimate a slope with a linear regression fit and thus estimate the Hurst exponent as $H = |\delta + 1|.$

One motivation for using the best basis to estimate the Hurst exponent in this fashion can be found by comparing this to the Maximum Modulus approach \cite{Mallat1992singularity} that is used when calculating scaling exponents with Continuous Wavelet Transforms (CWTs). In that context, the evaluation goes along paths ("snakes") defined by the most energetic wavelet coefficients. The decay of energy in the snakes along the scale is used to estimate regular scaling. The Jones {\it et al.} method insists on the “best” basis in a wavelet packet decomposition for related reasons. The best basis determined by minimizing the additive entropy measure results in the most disbalanced decomposition, in the sense that most of the coefficients are non-energetic and that most of the energy is contained in a few wavelet packet coefficients.

As described in more detail Section \ref{sec:Classification}, the blood protein spectra signals were divided into smaller sections, as in \cite{Dixon2023}, and each of these methods were used to estimate slopes for the sections to be used as features for classification.

\subsection{Feature Extraction}\label{sec:features}
A direct feature extraction method has been used in many studies to analyze the aforementioned ovarian cancer datasets \cite{R6}. This method uses the whole mass spectra to select discriminatory features with large differences in mean values between Cases and Controls characterized by Fisher's criterion,  
\begin{equation}\label{eqn-3}
    F = \frac{(\mu_{case} - \mu_{control})^2}{\sigma_{case}^2 + \sigma_{control}^2}
\end{equation}
where $\mu_{case}$ and $\mu_{control}$ are the arithmetic mean for the intensities at each mass-to-charge ratio in the mass spectra for the Case and Control groups, respectively, and $\sigma_{case}^2$ and $\sigma_{control}^2$ are the sample variances of the corresponding intensities over the mass-to-charge ratios in the mass spectra for the Case and Control groups. This measure is computed for all 15,153 mass-to-charge ratios in the mass spectra. The mass-to-charge ratios with the highest $F$ values are then selected for subsequent classification purposes. Prior studies indicate that using ten mass-to-charge ratios with the largest $F$ values can achieve good classification performance \cite{R6}. 

This study, however, uses an evolutionary spectra-based approach for deriving discriminatory features from the mass spectra. The evolutionary spectra-based method uses a moving data window to select features. That is, discriminatory features are selected from a data window while moving the data window along the mass spectra with a specific step size. Thus, this approach enables capturing more localized features in the mass spectra while still accounting for intensities of mass-to-charge-ratios. Spectral slopes of the wavelet spectra of those data windows are extracted as discriminatory descriptors. Depending on the step size used to move the data window along the mass spectra, those descriptors could be extracted from overlapping (or non-overlapping) data windows. This study uses overlapping data windows, aiming to extract as many discriminatory features as possible from the mass spectra.

Fisher's criterion (\ref{eqn-3}) is then used to select a set of the most informative data windows in the mass spectra that well discriminate the Cases and Controls. To compute the $F$ value for each data window, the mean and variance ($\mu$ and $\sigma^2$) of slopes of each data window corresponding to the Case and Control groups are used. Then, a set of data windows that correspond to the highest $F$ values are selected as the most informative discriminatory features.

\section{Ovarian Cancer Spectra Classification}\label{sec:Classification}
This section evaluates the performance of the proposed modalities in detecting ovarian cancer. We compare classification accuracy using the DWT, Wang {\it et al.}, and Jones {\it et al.} methods for the {\it Ovarian} 4-3-02 and {\it Ovarian} 8-7-02 datasets.

\subsection{Data Preparation}\label{sec:preprocess}

The following data preparation process was performed in MATLAB (computing scaling descriptors) and Python (train/test splitting and feature selection).

\begin{itemize}
    \item [(1)] Within a data window of length 1024, the wavelet packet tree is computed for each of the protein mass spectra within this window (for the DWT-based and Wang {\it et al.} methods, $J = 10$ levels of decomposition with the Haar wavelet were used; for the Jones {\it et al.} method, $J = 9$ levels with the Symmlet wavelet with four vanishing moments (SYMM4) were used).
    
    \item [(2)] The slopes are estimated for each method using the WPD coefficients obtained in Step (1). See \ref{sec:preprocess_details} for more details about specific steps taken for each method.

    \item [(3)] Wavelet-based features are selected in the following manner:
    \begin{itemize}
        \item [(a)] Steps (1) and (2) are repeated while shifting the data window by 500 data points along the protein mass spectra. To cover nearly all of the 15,153 mass-to-charge ratios, $\lfloor (15153-1024)/500 + 1 \rfloor = 29$ overlapping windows are required and cover 15,024 mass-to-charge ratios (i.e., resulting in 29 spectral slope features). The largest 129 mass-to-charge ratios are not included in the analysis. Two feature matrices are then formed using the estimated slopes, and they are denoted as $X_{N_1 \times 29}^{control}$ and $X_{N_2 \times 29}^{case}$, where $N_1$ and $N_2$ is the number of spectra in the Control and Case groups, respectively. 
        
        \item [(b)] To select most significant classifying features, Fisher's criterion ($F$) given in Equation (\ref{eqn-3}) is computed, and the $p$ features with the largest $F$ values from the feature matrices are then selected for classification purposes. This results in smaller feature matrices $ X_{N_1 \times p}^{control}$ and $X_{N_2 \times p}^{case}$ where $p \le 29$.

        \item [(c)] Response matrices are created by assigning labels 1 and 0 for Cases and Controls respectively ($y_{N_1 \times 1}^{control}$ and $y_{N_2 \times 1}^{case}$).
       \end{itemize}
    \end{itemize}

\subsection{Computing Scaling Descriptors}\label{sec:preprocess_details}

\begin{itemize}
\item [(1)]{\bf DWT-based method: }
The DWT-based wavelet spectra was calculated using the second node in each level of the WPD from Step (1), since these nodes are just the detail coefficients for the respective levels of the DWT. See Section \ref{sec:ss} for details about calculating slopes from wavelet spectra. 

The wavelet spectra may deviate from linearity at the coarsest and/or finest levels when using real data, in which case choosing a subset of levels will improve estimation of the spectral slope. In order to determine which levels to use for this study, four observations were randomly selected from the Case and Control groups each from each dataset, and then the wavelet spectra plot was examined for the chosen observations across each data window (i.e. $4 \times 2 \times 29$ plots per dataset). The randomly chosen observations for {\it Ovarian} 4-3-02 were: Control - 101, 130, 142, 172; Case - 10, 15, 19, 34. For {\it Ovarian} 8-7-02, they were: Control - 199, 200, 211, 223; Case - 5, 34, 107, 142. The spectra were assessed visually to determine what subset of levels for each group showed the least deviation. 

After examining all data windows, a subset of levels was chosen to give the best coverage across all windows for each group. A breakdown of the specific levels chosen for each group and dataset is given in Appendix \ref{sec:wang_Appendix}. Only the chosen levels were used for estimating slopes in Step (2).

\item [(2)]{\bf Wang et al. method: }
The Wang {\it et al.} method wavelet spectra was calculated using the detail nodes from the WPD from Step (1). See Appendix \ref{sec:wang_Appendix} for details about calculating slopes using the Wang {\it et al.} method. As in the DWT-based method, only a subset of the levels was used to estimate the slopes. The same technique was used with the same randomly chosen observations, and the chosen levels are listed in Appendix \ref{sec:wang_Appendix}.

\item [(3)] { \bf Jones et al. method: }
After calculating the WPD in Step (1), the best basis nodes were identified as described in Appendix \ref{sec:bb_Appendix} using the non-normalized Shannon entropy as the cost function. The absolute values of the best basis coefficients were placed in a single vector, $C$, and sorted, and an index vector from 1 to 1024, $N_I$, was generated. No values from $C$ were discarded. The slope of $\log C$ vs. $\log N_I$ was calculated. 

\end{itemize}

\subsection{Classification Models and Performance}\label{sec:performance}

First, the Wilcoxon rank-sum test was used to verify that the estimated slopes for the Case and Control groups from the DWT, Wang {\it et al.}, and Jones {\it et al.} methods were highly significantly different ($p < 0.001$). Since {\it Ovarian} 8-7-02 has $N_1 = 91$ Controls and $N_2 = 162$ Cases, 91 of the 162 Cases were randomly selected to match the number of Controls prior to randomly determining the training and test sets to avoid bias due to imbalanced group sizes. 

Next, classification performance was evaluated using scikit-learn \cite{scikit-learn} in Python with the following four algorithms and parameters: logistic regression (L2 penalty, C = 1.0 and limited-memory BFGS solver), support vector machine (radial basis function kernel and C = 1.0), K-nearest neighbors (5 neighbors, equal weights and standard Euclidean distance), and random forest (Gini impurity and 100 trees). 

\begin{figure}[!t]
	\centering
	\includegraphics[width=\linewidth]{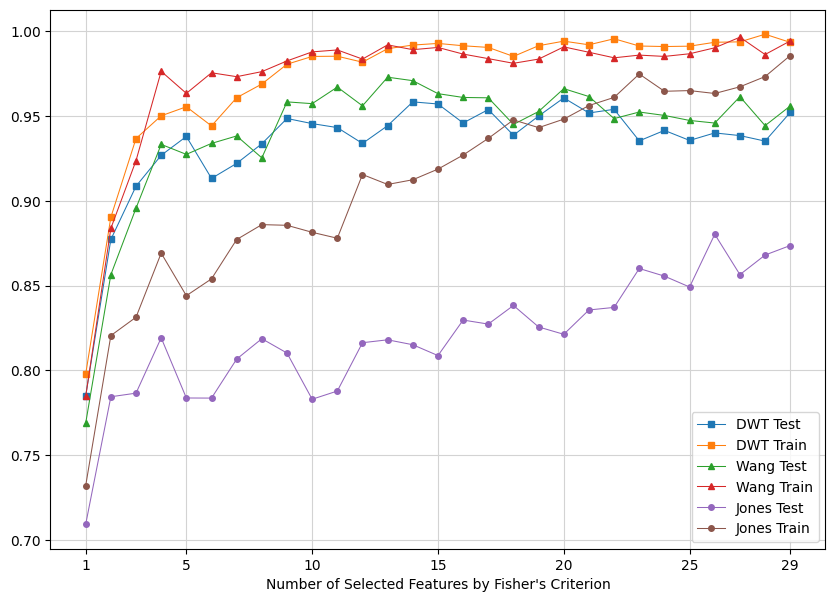}
	\caption{Average SVM training and test set accuracy over 1,000 repetitions over 1-29 predictors for the {\it Ovarian} 8-7-02 dataset}
	\label{fts_acc}
\end{figure}

For model fitting, 67\% of the samples were randomly selected and used for training; the samples in the remaining rows were used for testing. To determine the optimal number of classifying features, model training with randomly sampled data was repeated 1,000 times, and average accuracy over the repetitions was evaluated over different numbers of features sorted by Fisher's criterion values. The use of ten features for each algorithm was determined after observing that additional features did not offer significant accuracy improvements for the best-performing algorithms as indicated in Fig. \ref{fts_acc}. Next, using the ten most significant features identified for each method and dataset, the average test set classification accuracy was evaluated for 10,000 different training and testing splits. 

For the {\it Ovarian} 4-3-02 dataset, the classification accuracies achieved with the DWT and wavelet packet-based methods are similar, with the DWT-based method performing slightly better overall (Table \ref{tab-1}). For the {\it Ovarian} 8-7-02 dataset, the features generated by the Wang {\it et al.} method achieved slightly better classification accuracies than those from the DWT-based method. The features generated with the Jones {\it et al.} method resulted in lower classification accuracies in both datasets. Hypotheses for the three methods' performance, future research areas, and biomarkers that are captured by this study's generated features will be discussed in the following section. 

\begin{table*}[!t]
	\centering
	\begin{tabular}{|l|l|c|c|c|c|}
        \hline
	    {\bf Dataset} & {\bf Model} & {\bf DWT} & {\bf Wang et al.} & {\bf Jones et al.} \\
		\hline
	    \multirow{3}{*}{{\bf Ovarian} 4-3-02} 
           & {\bf Logistic Regression}  & 81.83 & 79.24 & 79.14 \\
           & {\bf Support Vector Machine}  & {\bf 83.07} & 80.38 & 81.05 \\
           & {\bf K-Nearest Neighbors}  & 82.48 & 80.12 & 79.05 \\
           & {\bf Random Forest}  & 81.93 & 77.65 & 80.54 \\
		\hline
	    \multirow{3}{*}{{\bf  Ovarian} 8-7-02} 
           & {\bf Logistic Regression}  & 94.90 & 96.48 & 79.17 \\
           & {\bf Support Vector Machine}  & 95.67 & {\bf 96.78} & 79.89 \\
           & {\bf K-Nearest Neighbors}  & 92.05 & 94.44 & 81.19 \\
           & {\bf Random Forest}  & 91.74 & 94.22 & 80.82 \\
		\hline
	\end{tabular}
	\vspace{.1cm}
	\caption{Average test set accuracy over 10,000 training and testing iterations with the ten most significant features with respect to Fisher's criterion for the DWT, \cite{Wang2006}, and \cite{Jones1996wavelet} methods.}
    \label{tab-1}
\end{table*}

\section{Discussion}\label{sec:Discussion}
 
This study underscores the significance of incorporating wavelet packets and accounting for scaling properties in protein mass spectra, as it contributes to enhanced ovarian cancer diagnosis efficacy. In the following subsections, we discuss the specific advantages of using wavelet-based methods and wavelet packets for ovarian cancer diagnosis in particular. We also review biomarkers that can be automatically captured by the proposed descriptors. 

\subsection{Wavelet Packets}\label{sec:disc-wave}

\begin{figure}[!t]
    \centering
    \includegraphics[width=\linewidth]{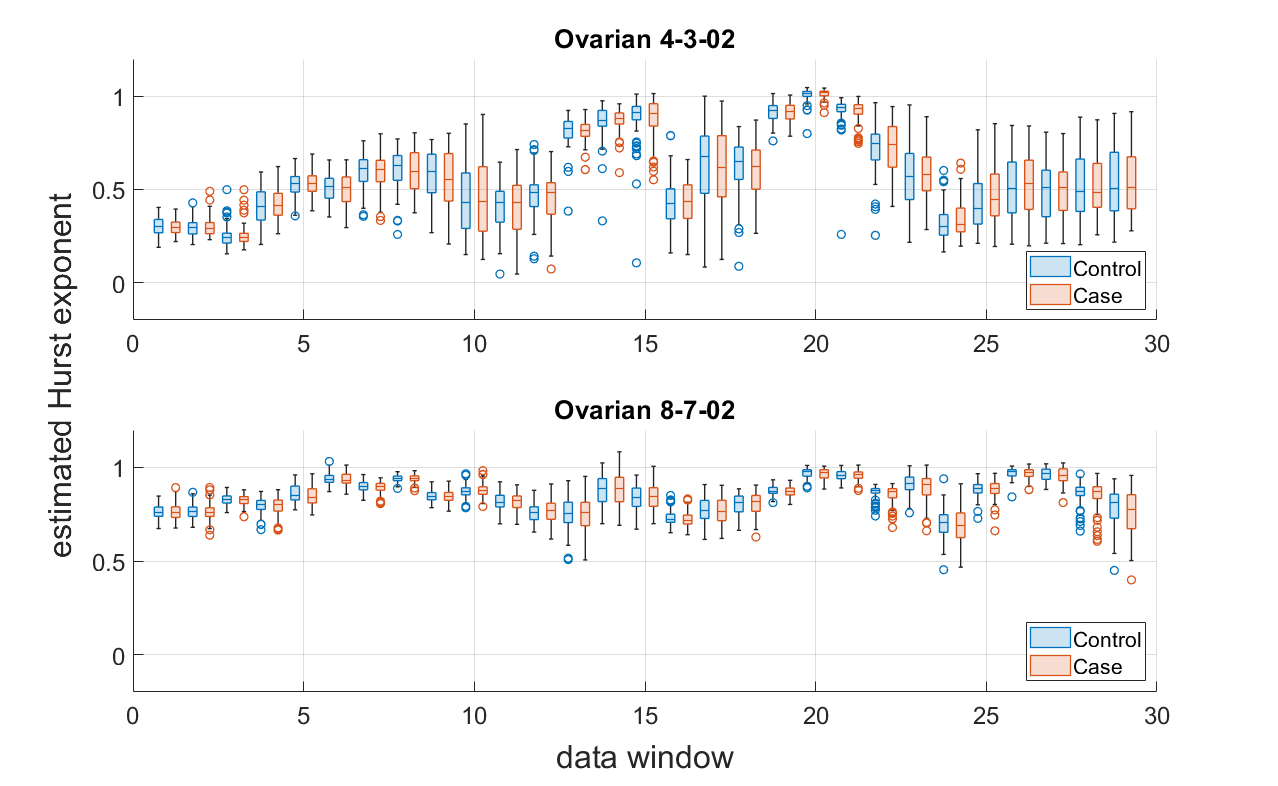}
    \caption{Distribution of the estimated Hurst exponent within each signal window for the {\it Ovarian} 4-3-02 and 8-7-02 datasets using the \cite{Wang2006} method. The overall range of the estimated $H$ values is smaller for the {\it Ovarian} 8-7-02 dataset.}
    \label{fig:compare-real-H-Wang}
\end{figure}

The method by Wang {\it et al.} may be well-suited for generating discriminatory features for datasets like {\it Ovarian} 8-7-02, in which the range of Hurst exponent estimates is small. As shown in Fig. \ref{fig:compare-real-H-Wang}, the Hurst exponents estimated using the Wang {\it et al.} method cover a smaller range for the {\it Ovarian} 8-7-02 dataset (mostly between 0.6 to 0.9) compared to the {\it Ovarian} 4-3-02 dataset (spanning the range from 0 to 1). In our simulations (see Table \ref{tab:wavelet_comparison}), although the DWT-based method showed the lowest bias in this range of Hurst exponents, it also had the largest variability. Conversely, while the method by Jones {\it et al.} had the lowest variability, it also had the worst bias. The Wang {\it et al.} method may provide accurate enough estimates of the self-similarity while also having low enough variability to separate signals with subtle differences in this smaller range. Since wavelet packets create a redundant representation of a signal, the method of Wang {\it et al.} takes advantage of this by calculating the average energy of all the detail coefficients at each level. In the DWT-based method, the number of available detail coefficients is halved at each level. This may explain why, between these two wavelet spectra-based methods, the Wang {\it et al.} method estimates the Hurst exponent with lower variability, since it uses a larger number of coefficients in the level-wise energy calculations.

The wavelet packet-based methods employed in this study inherit many of the advantages of standard wavelet-based methods, such as requiring minimal preprocessing of the data and offering the ability to capture co-expression of proteins. As described in \cite{Dixon2023}, previous studies which extract features from complex protein mass spectra employ preprocessing steps such as baseline correction, peak alignment, normalization, and denoising. The method by Wang {\it et al.} does not require any preprocessing steps beyond those which would already be employed for the DWT-based method (primarily, choosing levels that are used to estimate the rate of energy decay in the wavelet spectra). The former method is also simple to implement, since it is conceptually similar and merely extends the DWT-based method to the case where there are multiple sets of coefficients at each level. As for capturing co-expression: groups of biomarkers have been studied for their ability to diagnose ovarian cancer in combination. For example, \cite{Hasenburg2021biomarker} used the levels of five proteins as features in a logistic regression model. In such studies, the individual proteins are manually selected, whereas measurements of self-similarity automatically describe intricate patterns within a signal, including how signal intensities relate across different indices. Thus, wavelet packets can offer a more automated approach for capturing these complex relationships. Another advantage of wavelet packets is that the features generated with them do not require additional features in order to achieve high classification accuracy. For example, \cite{Dixon2023} demonstrated that DWT-based features led to an improvement in classification accuracy when combined with standard intensity measurements. However, models built exclusively with DWT-based features achieved less than 90\% accuracy even when up to ten features were included. 

\subsection{Biomarkers}\label{sec:disc-biomarkers}
Ovarian cancer, a complex disease with multiple potential biomarkers, has been the subject of extensive research for diagnostic purposes. Mass spectrometry is a widely used technique for identifying potential biomarkers, with the mass-to-charge (m/z) ratio being a critical parameter \cite{mass-spec}. In this study, we explored the capacity of the proposed scaling descriptors to identify and capture m/z ratios associated with key biomarkers. 

\begin{figure*}[!t]
	\centering
	\begin{subfigure}{.55\paperwidth}
		\includegraphics[width=.6\paperwidth]{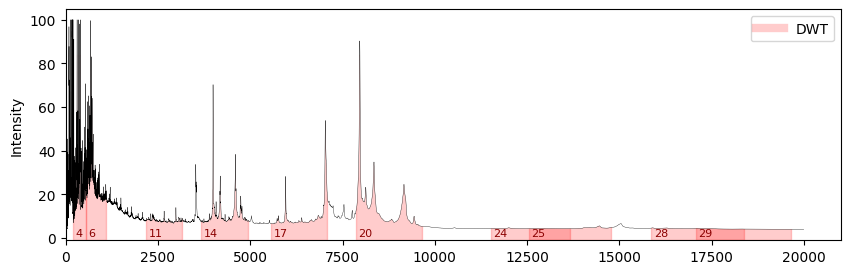}
	\end{subfigure}\\
	\begin{subfigure}{.55\paperwidth}
		\includegraphics[width=.6\paperwidth]{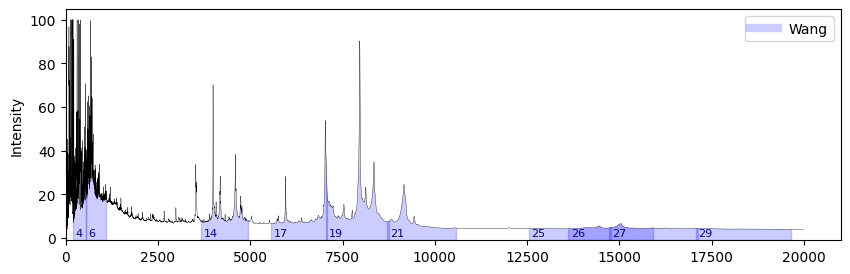}
	\end{subfigure}\\
	\begin{subfigure}{.55\paperwidth}
	        \includegraphics[width=.6\paperwidth]{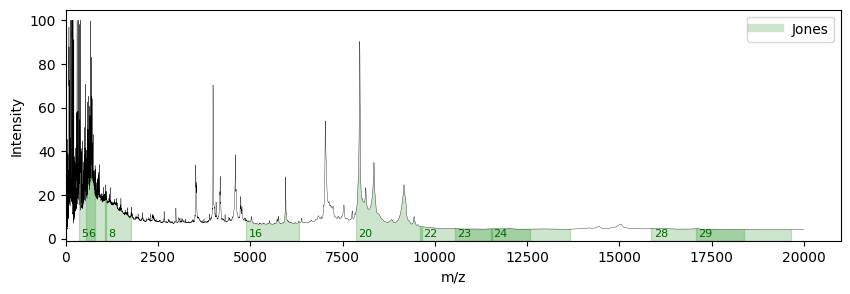}
         \end{subfigure}
	\caption{Protein mass spectra for an individual Case sample from {\it Ovarian} 8-7-02. Highlighted ranges of mass-to-charge ratios reflect the ranges for which spectral descriptors are determined to be significant (determined by Fisher's criterion) in classifying ovarian cancer using three wavelet-based methods, with integer labels within ranges corresponding to each of the 29 windows.}
    \label{mz_fts}
\end{figure*}

Fig. \ref{mz_fts} visually presents the m/z windows corresponding to the top ten most significant scaling descriptors, identified with Fisher's criterion, for the DWT, Wang {\it et al.} and Jones {\it et al.} methods. Integer labels in the highlighted windows correspond to each of the 29 windows used to cover the mass-to-charge ratios (e.g., "1" corresponds to the first window, with m/z indices from 1 to 1024, while "2" corresponds to the second window, with indices from 501 to 1524; see the "Data Preparation" section above). 

Of note, the scaling descriptors successfully capture m/z regions associated with several key biomarkers indicative of ovarian cancer presence. For example, transgelin 2, which is a potential biomarker, corresponds to an m/z value of 2345.19 in the protein mass spectra and is encompassed within window 11. Table \ref{tab-biomarkers} provides a summary of the biomarkers contained within these windows \cite{biomk-1, biomk-2, biomk-3, biomk-4}. In essence, the proposed scaling descriptors offer an automated means of identifying several previously recognized biomarkers, as well as new mass-to-charge ratios that hold promise as crucial biomarkers for ovarian cancer detection. This demonstrates the valuable potential of the scaling descriptors in advancing our understanding and diagnosis of ovarian cancer. 

\begin{table*}[!t]
    \centering
    \begin{tabular}{|c|c|c|l|l|l|}    
    \hline
        {\bf Window} & {\bf m/z Indices} & {\bf m/z Values} & 
        {\bf Methods} & {\bf Biomarkers} \\\hline 
        
        \multirow{2}{*}{6}	&	\multirow{2}{*}{2,501 -	3,524}	&	\multirow{2}{*}{543.93	-	1,080.39}	&	DWT, Wang et al.,  &  
           \multirow{2}{*}{Keratin 2a}  \\ 
        & & &Jones et al. &   \\            
           \hline
        
       \multirow{3}{*}{8}	&	\multirow{3}{*}{3,501 -	4,524}	&	\multirow{3}{*}{1,066.33 -	1,780.99}	&	\multirow{3}{*}{Jones et al.} &	
        Complement component 4A   \\
        & & & & preproprotein  \\
        & & & & Glycosyltransferase-like 1B  \\ 
        \hline
        
        \multirow{4}{*}{11}	&	\multirow{4}{*}{5,001 -	6,024}	&	\multirow{4}{*}{2,176.53 -	3,158.48}	&	\multirow{4}{*}{DWT}	& Transthyretin  \\ 
        & & & & Inter-\textalpha-trypsin inhibitor	\\ 
        & & & &  heavy chain H4	\\ 
        &&&& Transgelin 2  \\ \hline
        19	&	9,001 -	10,024	&	7,053.15 -	8,747.88	&	Wang et al.&	Apolipoprotein A2 \\ \hline
        20	&	9,501 -	10,524	&	7,858.69 -	9,642.52	&	DWT, Jones et al. & Apolipoprotein A2\\ \hline
                
    \end{tabular}
    \caption{Potential ovarian cancer biomarkers that coincide with feature windows from this study that were identified as significant by Fisher's criterion.}
    \label{tab-biomarkers}
\end{table*}

\subsection{Drawbacks}\label{sec:disc-drawbacks}

A drawback common to methods involving a “rolling window,” and in particular to all of the methods demonstrated here, is the subjective choice of the data window size, which could change or generate different features and thus influence the performance of the resulting models. A larger window size would allow the individual scaling descriptors to characterize the interplay between more protein expressions, but conversely could lead to an averaging effect that smooths out subtle patterns. On the other hand, a smaller window size may allow for finer characterization of localized patterns in the signals, but could quickly lead to an overabundance of features relative to the available samples. The length of the data windows is also related to the number of levels of DWT or WPD that can be carried out, which affects the number of points available for estimating the spectral slope in the DWT and Wang {\it et al.} methods. The authors in \cite{Dixon2023} found that 1024 points per window worked well in related studies, and so this choice of window size was adopted here.

\subsection{Classification Accuracy}\label{sec:disc-classificationacc}

\begin{figure}[!t]
\centering
\includegraphics[width=0.9\linewidth, height = 0.6\paperwidth]{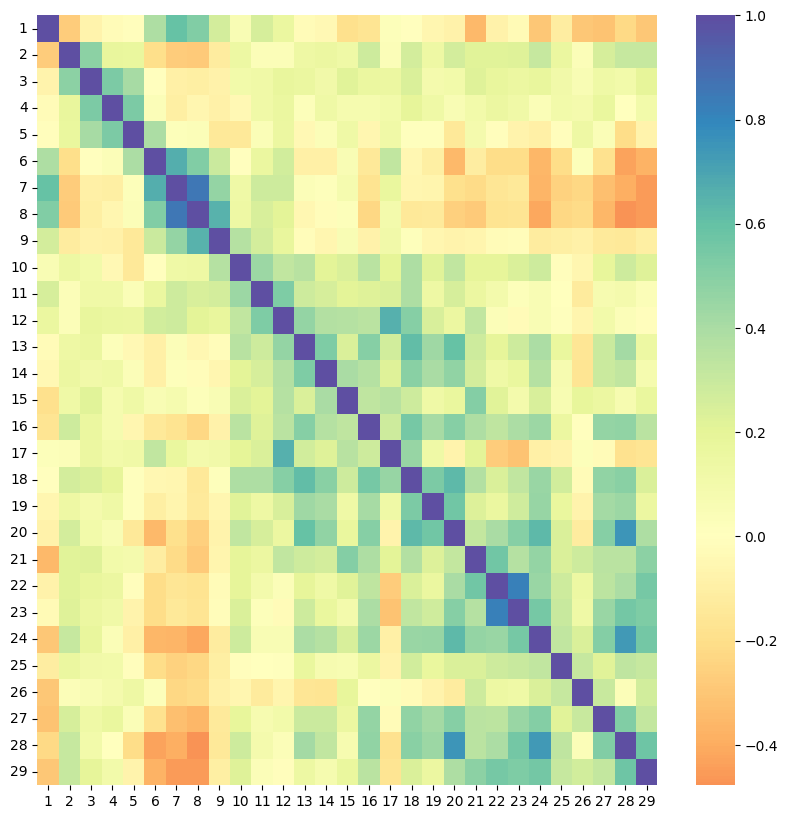}
    \caption{Correlation coefficient matrix for features from the Jones {\it et al.} method for {\it Ovarian} 8-7-02 data indicate that a significant number of feature pairs show moderate to strong correlation}. 
    \label{corr_jones}
\end{figure}

As stated in \cite{R39}, the {\it Ovarian} 8-7-02 dataset was collected with an upgraded instrument, a potential reason for the higher performance of the classification algorithms on this dataset as compared to {\it Ovarian} 4-3-02. The Jones {\it et al.} method yielded several highly correlated features, as shown in Fig. \ref{corr_jones}. This likely contributes to overfitting of the classifiers, evidenced by the divergence between testing and training accuracies for this method (see Fig. \ref{fts_acc}), thus leading to lower test classification accuracy than that of the DWT and Wang {\it et al.} methods.

As it is important to combine several instruments for cancer testing, this paper provides a rapid and robust quantitative measure to strengthen existing ovarian cancer diagnostic procedures. Although the accuracy rates could be argued to be relatively low for {\it Ovarian} 4-3-02, even classifiers that perform only slightly better than chance can improve diagnostic accuracy when added to a battery of other independent testing modalities.

\section{Conclusion}\label{sec:Conclusion}

This paper introduces a novel set of scaling descriptors rooted in the self-similarity of protein mass spectra and evaluates their performance in ovarian cancer diagnostics. The self-similar characteristics of data signals are examined using two wavelet packet-based techniques applied to protein mass spectra. We then explore the ability of these descriptors to improve ovarian cancer detection by using them as features in several classification algorithms.

This study has shown that wavelet packet-based descriptors using the Wang {\it et al.} method can improve classification accuracy over the DWT method, and also can achieve performance approaching that of existing methods that are based on a combination of both wavelet and direct mass-to-charge ratio-based descriptors \cite{Dixon2023}. For both datasets, classification accuracy falls within the range of 79\% to 97\% for the proposed wavelet packet-based features, and 82\% to 96\% for the DWT-based features. Since early detection of ovarian cancer significantly increases survival rates, any improvement in detection methods is highly significant. The proposed approach captures new mass-to-charge ratio ranges with minimal data preprocessing, while also encompassing mass-to-charge ratio ranges that encompass the majority of previously identified biomarkers.

However, it is essential to acknowledge certain limitations of the proposed method, which could be addressed in future research. Specifically, accounting for machine effects and measurement noise in modeling efforts is crucial, as these factors can adversely impact classification performance. This would involve exploring models with careful fine-tuning to mitigate the influence of such nuisance information. Notably, \cite{R44} discusses potential machine learning methods to address these challenges, providing a valuable avenue for future research. 

In the spirit of reproducible research, the software used in this paper is posted on \url{https://github.com/jihyun-byun/ovarian-cancer-classification}.

\section*{Acknowledgements}

This research was supported in part by the H.O. Hartley endowment at Texas A\&M University. The study sponsor had no involvement in the study design, collection, analysis and interpretation of the data; in the writing of the manuscript; and in the decision to submit the manuscript for publication.

\appendix

\section{Discrete Wavelet Transform}\label{sec:wt1_Appendix}

Suppose a data signal ($Y$) is a vector of size  $N \times 1$. The DWT of $Y$, denoted as  $\mathbf{d}$, is represented as
\begin{equation}\label{eq-12}
  \mathbf{d} = WY,
\end{equation}
where $W$ is an orthogonal matrix of size ${N \times N}$. The elements in $W$ are determined by selecting a particular wavelet basis, such as Haar, Daubechies, and Symmlet. Although $N$ can be arbitrary, it is usually selected to be a power of two, i.e., $N = 2^J$, $J \in \mathbb{Z}^+$ for the ease of calculating the DWT.

When $N$ is large, and calculating $\mathbf{d}$ as in (\ref{eq-12}) becomes computationally expensive, a fast algorithm based on filtering proposed by Mallat is used for computational efficiency. With this algorithm the DWT is obtained by performing a series of successive convolutions that involve a wavelet-specific low-pass filter $\mathbf{h}$ and its quadrature-mirror counterpart, high-pass filter $\mathbf{g}$. These repeated convolutions using the two filters accompanied with the operation of decimation (keeping every second coefficient of the convolution) generate a multiresolution representation of a signal, consisting of a smooth approximation coefficients, $c_{jk}$, and a hierarchy of detail coefficients $d_{jk}$ at different resolutions (indexed by a scale index $j$) and different locations within the same resolution, indexed by $k$. The convolutions with filters $\mathbf{h}$ and $\mathbf{g}$ are repeated until a desired decomposition level $j=J_0$ is reached ($1 \leq J_0 \leq J-1 $ and $J=\log_2 N$).

Thus, the vector $\mathbf{d}{}$ in (\ref{eq-12}) has a structure,
\begin{equation}\label{eq-12a}
  \mathbf{d} = (\mathbf{c}_{J_0}, \mathbf{d}_{J_0}, \dots, \mathbf{d}_{J-2}, \mathbf{d}_{J-1}),
\end{equation}
where $\mathbf{c}_{J_0}$ is a vector of coefficients corresponding to a smooth trend in signal, and $\mathbf{d}_{j}$ are detail coefficients at different resolutions $j$ where $J_0 \leq j \leq J-1.$  It is the logarithm of the variance of coefficients in $\mathbf{d}_j$ that is used to define wavelet spectra.

\section{Wavelet Packet Decomposition}\label{sec:wpt_Appendix}

The wavelet packet decomposition (WPD) extends the wavelet transform by analyzing the detail coefficients in addition to the approximation coefficients, thereby generating a tree of coefficients that may better characterize certain qualities of the signal.

The WPD of a $N \times 1$ signal $Y$ can be calculated as $\mathbf{d} = WY$. Here, $W$ is a $NJ \times NJ$ matrix that is determined by the choice of wavelet and number of levels of decomposition, $J$.

The wavelet packet coefficients may be calculated as follows. Let $\mathbf{h}$ and $\mathbf{g}$ be the low- and high-pass filters, respectively, determined by the choice of wavelet, and let $\mathcal{W}_0 = \psi (x)$ and $\mathcal{W}_1 = \phi (x)$ be the corresponding scaling and wavelet functions. Next, define a library of wavelet packet functions, $\{\mathcal{W}_n (x)\}$, as follows, which generates the entire wavelet packet tree or “table”:
\begin{align*}
    \mathcal{W}_{2n}(x) &= \sum_k h_k \sqrt{2} \mathcal{W}_n (2x - k), \\
    \mathcal{W}_{2n+1}(x) &= \sum_k g_k \sqrt{2} \mathcal{W}_n (2x - k), n = 0, 1, 2, \ldots \ .
\end{align*}

This library of functions is overcomplete, or redundant in a sense, and an arbitrary subset chosen from it may not have desirable properties (i.e. being an orthonormal and/or complete set). In order to choose an appropriate subset (an orthonormal basis), instead define the wavelet packet functions in terms of three parameters: $j$ - the scaling parameter (corresponding to the level of the tree); $n$ - the oscillation parameter (corresponding to the node within a level); and $k$ - the translation parameter (corresponding to the coefficient index within a node). Then, the wavelet packet functions can be written as follows:
\begin{equation}\label{Wjnk}
    \mathcal{W}_{j,n,k} = 2^{j/2} \mathcal{W}_n (2^j x - k), 
        (j, n, k) \in \mathbb{Z} \times \mathbb{N} \times \mathbb{Z},
\end{equation}
and any suitable (i.e. $\mathbb{L}_2$) function $f$ can now be represented as:
\begin{equation}
    f(x) = \sum_{(j, n) \in \mathbb{P}} {
        \sum_{k \in \mathbb{Z}} {
            w_{j,n,k} \mathcal{W}_{j,n,k}(x),
        }
    }
\end{equation}
where $\mathbf{w}_{J, 0}$ is conventionally the discrete dataset. 

Here, $(j, n) \in \mathbb{P}$ means that we must select nodes from the wavelet packet table that meet special criteria in order for them to form a valid orthonormal basis. Formally, as described in, e.g., \cite{Vidakovic1999statwavelets}, the dyadic intervals $\mathcal{I}_{j,n} = \{ [2^j n , 2^{j(n+1)}, \{j, n\} \in \mathbb{P} \in \mathbb{Z} \times \mathbb{N} ) \}$ must form a disjoint, countable covering of the interval $[0, \infty)$. Informally, if we draw an arbitrary vertical line down a wavelet packet table, then the line must intersect a selected node exactly once. More than one intersection results in redundancy, while no intersection results in non-completeness.

\section{Wang et al. method calculations}\label{sec:wang_Appendix}

For a signal of length $N = 2^J$, level $j = J, J-1, \ldots, J-j_0$ of the wavelet packet tree contains $2^{J-j-1}$ detail nodes (indexed by $k$) with $N / 2^{J-j}$ coefficients (indexed by $i$) in each node. The energy for a node is calculated similarly as in the DWT method, $E_{jk} = \sum_{i=1} ^ {N / (2^{J-j})} {c_{ijk}^{2}}$. Then, the energy for the level is the average of the detail node energies, $E_j = (1 / 2^{J-j-1}) \sum _{k=1} ^{2^{J-j-1}} {E_{jk}}$. This results in an energy $E_j$ for each level. The wavelet spectra is plotted and the spectral slope is estimated in the same way as the DWT method (described in Section \ref{sec:ss}). As described in \cite{Wang2006}, the Hurst exponent is estimated as $H = -(slope/2)$.

The following levels were used when estimating the slope with the DWT method: {\it Ovarian} 4-3-02 - windows 1-11, levels 7-10; windows 12-29, levels 6-9; {\it Ovarian} 8-7-02 - windows 1-11, levels 8-10; windows 12-29, levels 7-10. And with the Wang {\it et al.} method, the following levels were used: {\it Ovarian} 4-3-02 - windows 1-10, levels 7-10; windows 11-29, levels 6-9; {\it Ovarian} 8-7-02 - windows 1-15, levels 8-10; windows 16-29, levels 6-10.

\section{Best Basis Selection}\label{sec:bb_Appendix}

A wavelet packet decomposition can be seen as a library of orthonormal bases. The best basis for a signal, which minimizes an information cost functional, can be found efficiently by exploiting the natural binary tree structure of the decomposition. In this study, the non-normalized Shannon entropy is used as the information cost functional, which for a vector $\mathbf{x} = (x_1, \ldots, x_n)$ can be calculated as
\begin{equation}
    \mathcal{C}(\mathbf{x}) = - \sum_{i = 1}^{n} x_i^2 \log x_i^2 \text{.}
\end{equation}

The method from \cite{Coifman1992entropy} finds the best basis in a way that only needs to examine each node in a tree twice. The algorithm can be carried out as follows, adapted from \cite{Wickerhauser1996adapted}.

\begin{enumerate}
    \item The cost function is calculated for each node in the tree, then all nodes in the final (bottom) level, $j=0$, are “marked”. 
    \item Next, visit each node in the level above, $j=1$. If the cost of a parent node from level 1 is less than the sum of the costs of its child nodes in level 0, then mark the parent. \label{bb_steploop_start}
    \item Otherwise, assign the sum of the child nodes’ costs to the parent node. \label{bb_steploop_end}
    \item Proceed upwards in the tree, applying steps \ref{bb_steploop_start}-\ref{bb_steploop_end} accordingly. 
    \item After examining all nodes in the tree, the topmost marked nodes are kept and all nodes under a marked node are pruned from the tree. The remaining nodes constitute the best basis.
\end{enumerate}

\bibliographystyle{unsrtnat}
\bibliography{references}  

\end{document}